\documentclass[a4paper,12pt]{article}

\usepackage{amsmath}
\usepackage{color}
\usepackage{hyperref}
\usepackage{axodraw}
\usepackage{graphicx}

\usepackage{cancel}
\usepackage{slashed}
\usepackage{amssymb}
\newcommand{\kay}{\slashed k}
\newcommand{\pone}{\slashed p_1}
\newcommand{\ptwo}{\slashed p_2}
\newcommand{\mh}{m_H}
\newcommand{\mt}{m_t}

\begin{document}
\begin{center} 

{\large\bf Higgs production as a probe of anomalous top couplings }\\[3ex]
Debajyoti Choudhury and Pratishruti Saha\\

\vspace*{5pt}

\begin{footnotesize}
{\sl Department of Physics and Astrophysics, University of Delhi, 
Delhi 110 007, India.}
\end{footnotesize}

\end{center} 

\vspace*{40pt}
 
\begin{abstract} 
\noindent The LHC may be currently seeing the first hints of the Higgs
boson. The dominant production mode for the Higgs at the LHC involves
a top-quark loop. An accurate measurement of Higgs production
cross-sections and decay widths can thus be used to obtain limits on
anomalous top couplings. We find that such an exercise could
potentially yield constraints that are stronger than those derived
from low-energy observables as well as direct bounds expected from the
$t \bar t $ production process.

\vspace*{30pt}
\noindent
\texttt{PACS Nos:14.65.Ha,14.70.Dj,14.70.Bh,14.80.Bn} \\ 
\texttt{Key Words:top,anomalous,higgs}
\end{abstract}

\vspace*{40pt}


\section{Introduction}
\label{sec:intro}

The Higgs boson is one of the key components of the Standard Model
(SM). And it is, as yet, undiscovered. Over the years, several
experiments have looked for the Higgs and ruled out its existence over
certain mass ranges. The first important step in this direction was
taken by the LEP experiments which ruled out a Higgs mass
less than 114.4 GeV~\cite{Barate:2003sz}. The latest in this line are
the CMS and ATLAS experiments at the LHC. The ATLAS collaboration has 
ruled out~\cite{ATLAS_Dec2011} the mass ranges 112.9--115.5 GeV, 
131--238 GeV and 251--466 GeV for the Standard Model Higgs while the 
CMS collaboration has ruled out~\cite{CMS_Dec2011} the entire range 
of 127--600 GeV. The Tevatron too has ruled out a subset of this
range~\cite{CDFandD0:2011aa}. However, that is not all. Lately, both
the CMS and the ATLAS~\cite{ATLAS_Dec2011,CMS_Dec2011} have observed the
first traces of what may be a signal for the SM Higgs. This excess is
seen the region 124--126 GeV and, although the statistics is
insufficient for a discovery to be claimed immediately, particle
physicists across the globe are 
enthused~\cite{pheno_refs_1,pheno_refs_2,pheno_refs_3} by the the fact
that the excess is seen in multiple channels and in the region
favoured by electroweak precision tests~\cite{LEPEWWG}.

If and when the Higgs is discovered, questions regarding its origin
and properties will have to be addressed. Is the Higgs only a single,
neutral CP-even scalar or does it have charged/CP-odd partners?  Is
the Higgs really a fundamental particle or is it a composite with
a dynamic origin?  The answers to some of these questions may involve
physics beyond the Standard Model and will be revealed by studying the
properties of the particle that is discovered. 
Two issues need to be borne in mind. Given that the SM,
necessarily, can only be an effective theory, what prevents the SM
Higgs from acquiring a large radiative correction to its mass? In
other words, the stabilization of the Higgs mass (or, equivalently,
the solution of the hierarchy problem) requires some mechanism beyond
the SM. In a related vein, a SM Higgs mass of $\sim 125$ GeV runs
afoul of the constraints from vacuum stability which, in turn, 
demands that there be some new physics at not too large a mass
scale.

Further, with the Higgs being responsible for the generation of masses
in the Standard Model and the top quark being the most massive
particle in the model, it is very likely that the top and the Higgs
sectors are closely intertwined. Probing one sector could reveal any
new physics in the other. This is explicitly borne out by a large
class of models that go beyond the SM in trying to explain
electroweak symmetry breaking~\cite{DyEWSB}. Of particular interest
in this context are mechanisms for dynamic breaking of electroweak
symmetry through the formation of $t\bar t$
condensates~\cite{tt_condensate}. Apart from these, many
other models could lead to large anomalous couplings of the top. A
partial list of examples wherein heavy fermions may play a role
would include Little Higgs models~\cite{lh0} or models with
extra space-time dimensions~\cite{acd_ued,ued_others,Barbieri_ed}. 
Similarly, if the SM is augmented by colour-triplet or colour-sextet 
scalars that have Yukawa couplings with the top-quark~\cite{diquark}, 
integrating out the former could also result in such eventualities.

The exact effects of such an extended sector on low-energy
observables would, understandably, depend on the details of the
model. However, on very general grounds, the act of integrating out
the heavy fields would result in higher-dimensional operators in the
low-energy effective theory~\cite{eff_terms}.
The form and magnitude of such operators
would depend on the nature and the sizes of the 
couplings that the SM fields under consideration have with those that 
have been integrated out. As we have argued above, the larger coupling 
of the top with the electroweak symmetry breaking sector is expected to 
play a defining role in such situations. Further, from a 
phenomenological point of view, the high threshold for top production
has meant that its couplings are still not well measured and can yet
accommodate significant deviations from the SM.

In this paper, we consider Higgs production and decay as probes of
anomalous couplings of the top quark. At the LHC, the primary mode for
Higgs production is through gluon fusion. As the latter is dominated
by the top loop, only those anomalous couplings of the top that lead
to any deviations in the $t t H$ and/or the $t t g$ vertices
are expected to modify the production rate. A modification of the 
$t t H$ vertex occurs even at the tree-level in theories with extra
Higgs fields and has already been explored extensively, for example,
in the context of supersymmetric theories. Here, we concentrate,
instead, on the top-gluon vertex, arguing that the limits already
obtained in the literature~\cite{previous,T_odd_corr,Choudhury:2009wd,
Rindani:1999gd,Atwood:1991ka,Rizzo:1994tu,Martinez:1996cy} still allow 
for substantial deviation of the Higgs production cross-section from 
its value within the SM.

Similarly, the Higgs decay width into a pair of photons can be
modified by invoking an anomalous $H W W$ vertex on the one hand,
and anomalous $W W \gamma$ or $t t \gamma$ vertices on the
other. Electroweak symmetry, though, relates the first of these to
the $H Z Z$ vertex, and the absence of any
deviation~\cite{ATLAS_Dec2011,CMS_Dec2011}
in either the already measured $Z Z^* \to 4 \ell$
channel or in the $W W^*$ channel strongly constrains any 
large modification of this vertex. Enhancement in this vertex is 
also constrained by the non-observation of a Higgs signal at the 
Tevatron~\cite{CDFandD0:2011aa}. Further, there are strong constraints 
from the LEP experiments on the modification of the $W W \gamma$ 
vertex. We shall, thus, limit ourselves to a discussion of the 
$t t \gamma$ vertex in the context of the Higgs decay.


\section{Analytic and Numerical Results}

\subsection{Anomalous \texorpdfstring{$t t g$}{top-gluon} couplings}
\label{sec:gluon}

The simplest gauge-invariant modification to the top-gluon 
vertex can be wrought by augmenting the Standard Model Lagrangian
by an effective operator of the form
\begin{equation}
{\cal L} \, \ni \, 
g_s\, {\cal A} \, (\bar Q_L \sigma_{\mu\nu} T^a t_R) \, F^{\mu\nu}_a \, \widetilde \phi 
\mspace{5mu} + \mspace{5mu} h.c. 
\label{eq:gg_lagrangian_1}
\end{equation}
where $Q_L$ (containing $t_L$) and the scalar field $\phi$ are the
usual SM doublets and $\widetilde \phi = -i \sigma_2 \phi^*$. The
constant ${\cal A} \sim {\cal O}(\Lambda^{-2})$ where $\Lambda$
represents the cut-off scale for the effective Lagrangian. Indeed,
the above is the only new Lorentz structure available as long as
one restricts to dim-6 operators. Any other modification can only be in 
terms of a momentum (form-factor) dependence of either ${\cal A}$ or 
even the canonical vector current. As this entails further suppression 
in $\Lambda^{-1}$, we shall neglect such behaviour.

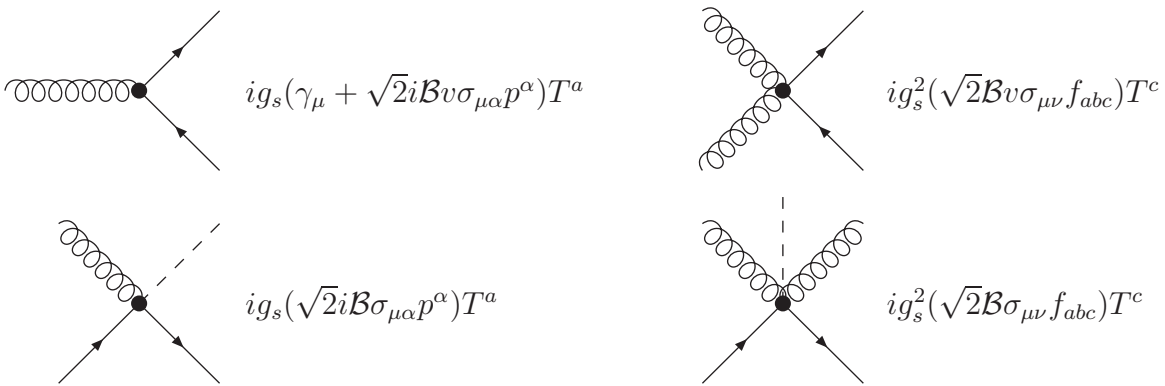
\begin{figure}[!htbp]
\vspace*{50pt}
\begin{picture}(155,120)(0,0)

\Gluon(10,140)(60,140){4}{7.5}
\ArrowLine(60,140)(90,170)
\ArrowLine(90,110)(60,140)
\Vertex(60,140){3}
    \Text(100,140)[l]{$ig_s ( \gamma_{\mu} + \sqrt{2} i {\cal B} v \sigma_{\mu\alpha} p^{\alpha} ) T^a$}

\Gluon(270,110)(300,140){4}{6.5}
\Gluon(270,170)(300,140){4}{6.5}
\ArrowLine(300,140)(330,170)
\ArrowLine(330,110)(300,140)
\Vertex(300,140){3}
    \Text(340,140)[l]{$ig_s^2 ( \sqrt{2} {\cal B} v \sigma_{\mu\nu} f_{abc} ) T^c$}

\Gluon(30,90)(60,60){4}{6.5}
\DashLine(60,60)(90,90){5}
\ArrowLine(30,30)(60,60)
\ArrowLine(60,60)(90,30)
\Vertex(60,60){3}
    \Text(100,60)[l]{$ig_s ( \sqrt{2} i {\cal B} \sigma_{\mu\alpha} p^{\alpha} ) T^a$}

\Gluon(270,90)(300,60){4}{6.5}
\Gluon(300,60)(330,90){4}{6.5}
\DashLine(300,60)(300,100){5}
\ArrowLine(270,30)(300,60)
\ArrowLine(300,60)(330,30)
\Vertex(300,60){3}
    \Text(340,60)[l]{$ig_s^2 ( \sqrt{2} {\cal B} \sigma_{\mu\nu} f_{abc} ) T^c$ }

\end{picture}
\vspace*{-30pt}
\caption{\em New Feynman rules.
$p^{\alpha}$ is the momentum of the incoming gluon. 
${\cal B} = Re({\cal A}) + i Im({\cal A}) \, \gamma_5$.}
\label{fig:ttg_vertices}
\end{figure}

After the breaking of electroweak symmetry, 
eq.(\ref{eq:gg_lagrangian_1}) gives
\begin{equation}
{\cal L} \, \ni \, 
g_s \, {\cal A} \, (\bar t_L \sigma_{\mu\nu} T^a t_R) \, F^{\mu\nu}_a \, \frac{(H + v)}{\sqrt{2}} 
\mspace{5mu} + \mspace{5mu} h.c.
\label{eq:gg_lagrangian_2}
\end{equation}
With the inclusion of this term, the $ttg$ gluon vertex gets modified. 
In addition, interaction terms involving $ttgg$, $ttgH$ and $ttggH$ 
are generated. The corresponding Feynman rules are given in 
Fig.\ref{fig:ttg_vertices}.

The terms proportional to $v$ correspond to anomalous 
chromomagnetic and chromoelectric dipole moments 
for the top and are often parameterized as 
\begin{equation}
\dfrac{g_s}{\Lambda} \, \bar t \, \sigma_{\mu\nu} T^a (\rho + i\rho' \gamma_5) t \, F^{\mu\nu}_a \, ,
\label{eq:gg_lagrangian_3}
\end{equation}
where, $\rho,\rho' =  +1, -1, 0$ and $\Lambda$ represents the energy 
scale of the new physics that may lead to such 
operators\footnote{In the Standard Model, 
  $\rho/\Lambda \sim {\cal O}(\alpha_s/\pi m_t)$ at the
  one-loop level. The coefficient of the $CP$-violating term,
  $\rho'/\Lambda$, is non-zero only at the three-loop level.}.
Equations \ref{eq:gg_lagrangian_2} and \ref{eq:gg_lagrangian_3} are,
thus, related by
\begin{equation}
\dfrac{\rho}{\Lambda} = \dfrac{v}{\sqrt{2}} \,Re({\cal A}) \ , \qquad
\dfrac{\rho'}{\Lambda} = \dfrac{v}{\sqrt{2}} \,Im({\cal A}) \ .
\label{eq:A_and_lambda}
\end{equation}
Clearly, the terms of eq.(\ref{eq:gg_lagrangian_3}) would 
modify top production rates from processes such as 
$p p, p \bar p \to t \bar t$ and $e^+ e^- \to t \bar t g$, and 
these have been used in the past to impose constraints on the couplings
\cite{Choudhury:2009wd,Rizzo:1994tu}.

A few comments are necessary at this stage. It might seem, 
at first sight, that we could have started with 
eq.(\ref{eq:gg_lagrangian_3}) rather than invoking 
eq.(\ref{eq:gg_lagrangian_1}) for the former is invariant under 
$SU(3)_C \times U(1)_{em}$. However, the lack of invariance under the 
full electroweak symmetry has consequences that we discuss below.
Also note that, although $\Lambda$ appears to be an arbitrary 
parameter, for the effective field theory formalism to be applicable,
$\Lambda$ must be greater than any other mass scale in the theory.
For example, Ref.~\cite{hasenfratz} suggests $\Lambda > 2\mt$.
Following this, we may parameterize
$\Lambda \equiv \zeta \mt$ where $\zeta > 2$. In other words, 
sensitivity to $\Lambda$ may be translated to sensitivity to the 
dimensionless parameter $\zeta$. We shall return to this discussion 
once again in Sec.~\ref{sec:summary}.

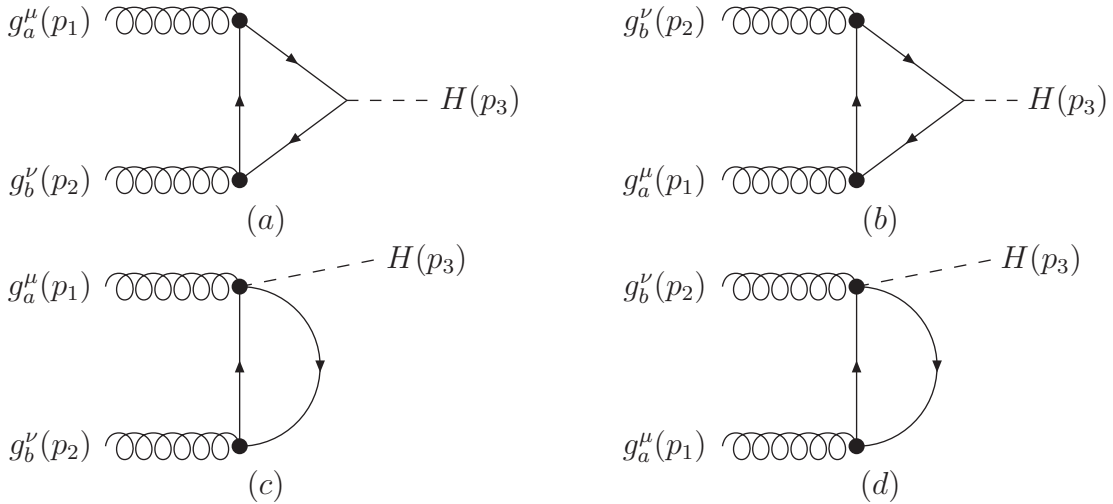
\begin{figure}[!htbp]
\vspace*{50pt}
\begin{picture}(155,120)(-75,0)

\Gluon(10, 100)(60, 100){5}{6.5}
    \Text(5, 100)[r]{$g_b^\nu(p_2)$}
\Gluon(10, 160)(60, 160){5}{6.5}
    \Text(5, 160)[r]{$g_a^\mu(p_1)$}
\ArrowLine(60,100)(60,160)
\ArrowLine(60,160)(100,130)
\ArrowLine(100,130)(60,100)
\DashLine(100,130)(130,130){5}
    \Text(135, 130)[l]{$H(p_3)$}
\Vertex(60,100){3}
\Vertex(60,160){3}
    \Text(70, -15)[c]{$(c)$}

\Gluon(240, 100)(290, 100){5}{6.5}
    \Text(235, 100)[r]{$g_a^\mu(p_1)$}
\Gluon(240, 160)(290, 160){5}{6.5}
    \Text(235, 160)[r]{$g_b^\nu(p_2)$}
\ArrowLine(290,100)(290,160)
\ArrowLine(290,160)(330,130)
\ArrowLine(330,130)(290,100)
\DashLine(330,130)(350,130){5}
    \Text(355, 130)[l]{$H(p_3)$}
\Vertex(290,100){3}
\Vertex(290,160){3}
    \Text(300, -15)[c]{$(d)$}

\Gluon(10, 0)(60, 0){5}{6.5}
    \Text(5, 0)[r]{$g_b^\nu(p_2)$}
\Gluon(10, 60)(60, 60){5}{6.5}
    \Text(5, 60)[r]{$g_a^\mu(p_1)$}
\ArrowLine(60,0)(60,60)
\ArrowArcn(60,30)(30,90,270)
\DashLine(60,60)(110,70){5}
    \Text(115, 70)[l]{$H(p_3)$}
\Vertex(60,0){3}
\Vertex(60,60){3}
    \Text(70, 85)[c]{$(a)$}

\Gluon(240, 0)(290, 0){5}{6.5}
    \Text(235, 0)[r]{$g_a^\mu(p_1)$}
\Gluon(240, 60)(290, 60){5}{6.5}
    \Text(235, 60)[r]{$g_b^\nu(p_2)$}
\ArrowLine(290,0)(290,60)
\ArrowArcn(290,30)(30,90,270)
\DashLine(290,60)(340,70){5}
    \Text(345,70)[l]{$H(p_3)$}
\Vertex(290,0){3}
\Vertex(290,60){3}
    \Text(300, 85)[c]{$(b)$}

\end{picture}
\vspace*{15pt}
\caption{\em LO amplitudes for $gg \to H$. $(a)$ and $(b)$ are 
analogous to the corresponding SM amplitudes with just the vertex 
factor being modified, whereas $(c)$ and $(d)$ two arise only for 
${\cal A} \neq 0$.}
\label{fig:ttg_diagrams}
\end{figure}

With the addition of the operator of 
eq.(\ref{eq:gg_lagrangian_2}) to the Lagrangian,
the lowest order (LO) amplitudes that contribute to $gg \to H$ are 
those shown in Fig.\ref{fig:ttg_diagrams}. Note that two of the 
diagrams, viz. Fig.\ref{fig:ttg_diagrams}$(c,d)$, would not arise if 
we had started with the operator of eq.(\ref{eq:gg_lagrangian_3}) instead.

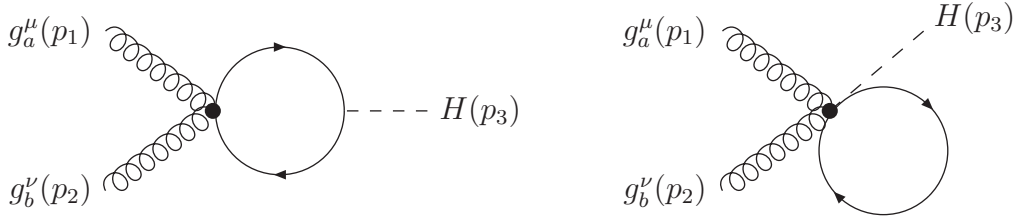
\begin{figure}[!htbp]
\vspace*{-40pt}
\begin{picture}(155,120)(-75,0)

\Gluon(10,0)(50,30){4}{7.5}
    \Text(5,0)[r]{$g_b^\nu(p_2)$}
\Gluon(10,60)(50,30){4}{7.5}
    \Text(5, 60)[r]{$g_a^\mu(p_1)$}
\ArrowArcn(75,30)(24,180,0)
\ArrowArcn(75,30)(24,0,180)
\DashLine(100,30)(130,30){5}
    \Text(135, 30)[l]{$H(p_3)$}
\Vertex(50,30){3}

\Gluon(240,0)(280,30){4}{7.5}
    \Text(235,0)[r]{$g_b^\nu(p_2)$}
\Gluon(240,60)(280,30){4}{7.5}
    \Text(235,60)[r]{$g_a^\mu(p_1)$}
\ArrowArcn(300,15)(24,135,-45)
\ArrowArcn(300,15)(24,-45,135)
\DashLine(280,30)(315,60){5}
    \Text(320,65)[l]{$H(p_3)$}
\Vertex(280,30){3}

\end{picture}
\vspace*{15pt}
\caption{\em Additional LO contributions to $gg \to H$ due to 
eq.(\ref{eq:gg_lagrangian_1}). These, however, are identically zero.}
\label{fig:extra_diagrams}
\end{figure}

Two additional diagrams (see Fig.\ref{fig:extra_diagrams}) also 
arise, but each of these can be seen to vanish identically being 
proportional to the trace of a single Gell-Mann matrix.
In other words, whereas the vertex structure stipulates that the 
two gluons need to be in a colour-antisymmetric state, for the process
under consideration, they must be in a singlet state. Indeed, both
these vertices can contribute only to the NLO processes, e.g., 
$g g \to H + g$.

The Standard Model amplitude for $gg \to H$ is finite. 
Indeed, this has to be so as such a term does not exist at the 
tree-level and hence no counterterms can be written. 
The introduction of the anomalous term changes matters considerably. 
Owing to its non-renormalizable nature, it could beget divergent 
quantum corrections to terms that were absent at the tree-level. 
The ggH vertex is one example of a term that could receive a divergent 
correction, but it is not the only one. Additional divergences could be 
generated at every higher order of perturbation theory. 
At the one-loop level though, the situation is under control. 
Consider, for example, Fig.2(a, b). With the higher dimensional nature 
of the anomalous coupling manifesting itself only in terms of the 
external momenta, the naive divergence of the loop remains unaltered 
from its SM counterpart. The situation would change for the worse if 
the anomalous coupling appeared at an internal vertex. This, however, 
can happen only at higher loops. Presently, we ignore this aspect as 
the theory under consideration is an effective one and such questions 
are meaningful only in the context of an ultraviolet completion.

Formally, the amplitudes corresponding to 
Fig.\ref{fig:ttg_diagrams}($a,b$) appear to be
linearly divergent, and, hence, a naive use of Feynman parameterization
and subsequent shifting of variables is fraught with danger, as it
might introduce non-trivial boundary terms. 
However, once these two amplitudes are added, the linearly divergent 
piece cancels exactly, leaving behind only a logarithmic divergence.
If we were to start with the Lagrangian of 
eq.(\ref{eq:gg_lagrangian_3}), the resultant amplitude would, 
typically, be proportional to $(\rho / \Lambda) \, \ln (\Lambda / M)$, 
where $M$ denotes some combination of the mass scales present in the 
theory, viz $\mt$, $\mh$. Such a structure is only to be expected in a 
non-renormalizable effective theory with a finite cut-off scale. 
Note that the term formally vanishes as the cutoff $\Lambda \to \infty$.
For the diagrams of Fig.\ref{fig:ttg_diagrams}$(c,d)$, once again,
the apparent quadratic divergence gets cancelled leaving behind a
only a logarithmic divergence.
Once the amplitudes from the two sets of diagrams 
(i.e. Fig.\ref{fig:ttg_diagrams}$(a,b)$ and 
Fig.\ref{fig:ttg_diagrams}$(c,d)$) are 
added, the divergences cancel exactly, leaving behind 
only a finite residue\footnote{Note, though, that the calculation of the 
finite residue has to be done with a gauge-invariant prescription. 
In other words, the individual logarithmic divergences need to be regularized 
in a gauge invariant manner, such as dimensional regularization.}.
A particular consequence of the inclusion of 
Fig.\ref{fig:ttg_diagrams}$(c,d)$ is that, for a given ${\cal A}$, 
the amplitude now is smaller (on account of the large logarithm 
vanishing) and, hence, the limits obtained on ${\cal A}$ are more 
conservative.

\begin{figure}[!htbp]
\begin{center}
\includegraphics[scale=0.7]{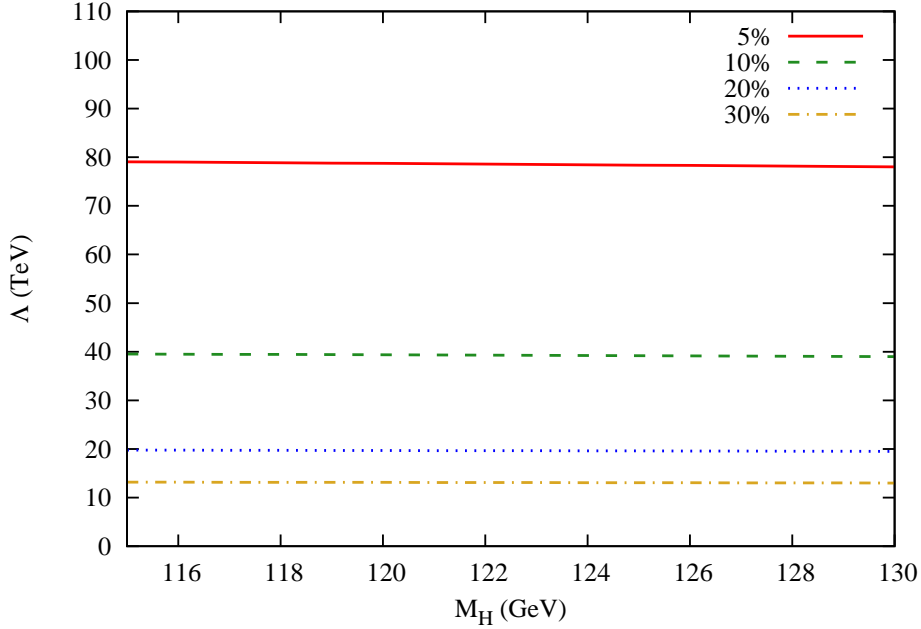}
\caption{\em Limits on $\Lambda$ (with $\rho = \pm 1$) obtained by 
imposing the restriction that the $\sigma(gg \to H)$ remain within 
m\% of its SM value. Results are plotted for m = 5, 10, 20, 30. 
In each case, the region below the curve is ruled out.}
\label{fig:gg_limits}
\end{center}
\end{figure}

Using this amplitude (the full expressions are given in the Appendix), 
one may then obtain the Higgs production 
cross-section as a function of ${\cal A}$. It is interesting to note 
that, as long as $Im({\cal A}) \neq 0$, a non-zero $CP$-violating 
coupling of the gluons to the Higgs is generated as well. 
This, of course, is not unexpected given the structure of the couplings.
However, in keeping with the spirit of effective field theories, 
we restrict ourselves to ${\cal O(A)}$ in our computation of the 
cross-section. Taking into account the ${\cal O}({\cal A}^2)$ terms 
would have necessitated the inclusion of ${\cal O}({\cal A}^2)$ terms in 
the Lagrangian (eq.(\ref{eq:gg_lagrangian_1})) itself.
Due to this, $Im({\cal A})$ does not contribute and 
consequently, constraints can be placed only on $Re({\cal A})$, or 
equivalently, on $\rho/\Lambda$. The allowed range for $\Lambda/\rho$ 
for different $m_H$ is shown in Fig.\ref{fig:gg_limits}. 
Allowing for a 30\% deviation in the cross-section, one finds that 
the region below the yellow (dot-dashed) line is ruled out. 
If one reduces the allowed deviation, a larger range of $\Lambda$
gets ruled out. This is the region below the blue (dotted) line for a 
deviation of 20\%, the green (dashed) line for 10\% and the red (solid) 
line for 5\%. As expected, greater accuracy in the measurement of the 
Higgs production cross-section leads to more stringent limits on 
$\Lambda$. We display the results for Higgs masses in the range 
115--130 GeV. In this range, the limits have only a very mild dependence 
on $m_H$.


\subsection{Anomalous \texorpdfstring{$t t \gamma$}{top-photon} couplings}
\label{sec:photon}

For a low mass Higgs that is favoured by current data, the most
promising discovery mode is $H \to \gamma\gamma$, and we now turn to
this channel. It could be argued that the signal strength for $p p \to
H \to \gamma \gamma$ may receive anomalous contributions at both the
production and the decay vertices and that unravelling the two is
impossible. Fortunately, though, excesses have been reported in 
$H \to Z Z^* \to 4 \ell$ channel as well and even the 
$H \to W W^*$ channel is being pursued assiduously.
Thus, observables such as
$\Gamma(H \to \gamma\gamma) / \Gamma(H \to Z Z^*)$ are likely to be 
well-measured. With the $HZZ$ and $HWW$ vertices being driven by tree 
level couplings~\footnote{These couplings are also experimentally 
  constrained by the Tevatron experiments~\cite{CDFandD0:2011aa}.}, 
the partial width $\Gamma(H \to \gamma\gamma)$ would be measurable and, 
thus, would constitute a very good probe for the effective 
$H \gamma\gamma$ vertex.

Analogous to eq.(\ref{eq:gg_lagrangian_2}), an effective operator for 
the $t t \gamma$ vertex can be written as
\begin{equation}
{\cal L} \, \ni \, 
e Q_t \, {\cal A}' \, (\bar t_L\sigma_{\mu\nu}t_R) \, F^{\mu\nu} \, \dfrac{(H + v)}{\sqrt{2}} 
\mspace{5mu} + \mspace{5mu} h.c., 
\label{eq:photon_lagrangian_2}
\end{equation}
where $Q_t$ is the charge of the top quark in units of $e$.
Once again, this may be written in terms of anomalous magnetic and 
electric dipole moments of the top quark.
\begin{equation}
\dfrac{\eta}{\Lambda'} = \dfrac{v}{\sqrt{2}} \, Re({\cal A'}) \ , \qquad
\dfrac{\eta'}{\Lambda'} = \dfrac{v}{\sqrt{2}} \,Im({\cal A'})
\label{eq:Apr_and_lambdapr}
\end{equation}
with $\eta,\eta' = +1, -1, 0$ and $\Lambda'$ being the relevant new 
physics scale. Note that the arguments in Sec.\ref{sec:gluon} 
pertaining to the lower limit on $\Lambda$, are also applicable for 
$\Lambda'$.

\begin{figure}[!htbp]
\begin{center}
\includegraphics[scale=0.7]{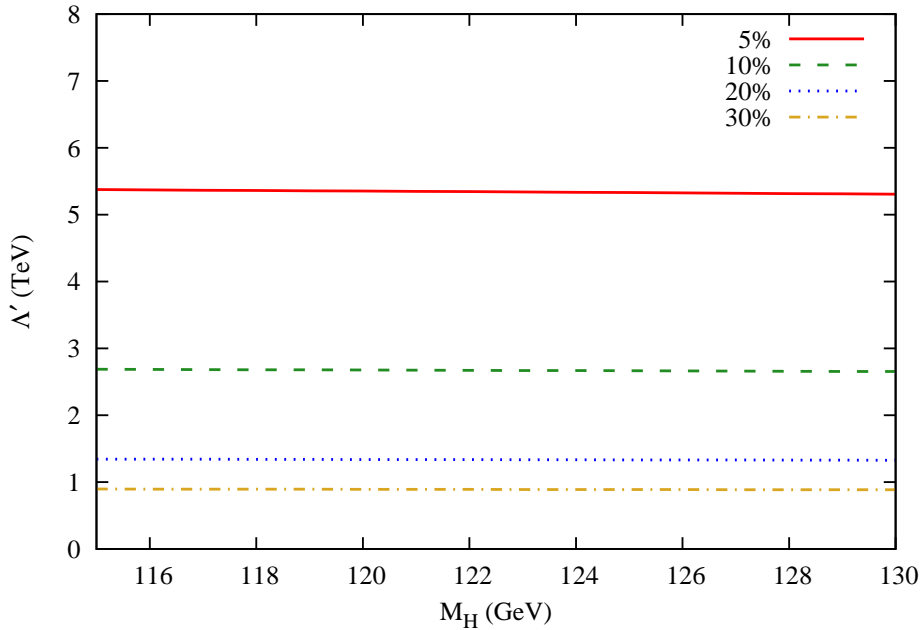}
\caption{\em Limits on $\Lambda'$ (with $\eta = \pm 1$) obtained by 
imposing the restriction that the $\Gamma(H \to \gamma\gamma)$ remain 
within m\% of its SM value. Results are plotted for m = 5, 10, 20, 30. 
In each case, the region below the curve is ruled out.}
\label{fig:photon_limits}
\end{center}
\end{figure}

The calculation of the top-loop contribution to 
$\Gamma(H \to \gamma\gamma)$ is akin to that for 
$\Gamma(H \to g g)$, except that there is no analogue of 
Fig.\ref{fig:extra_diagrams} (which, in any case, vanished identically).
Once again, restricting to ${\cal O}({\cal A}')$ 
results in limits being obtained only on $\eta/\Lambda'$. 
These are shown in Fig.\ref{fig:photon_limits}. $\Lambda'$ 
values below the yellow (dot-dashed) lead to a deviation of 30\% or 
more in $\Gamma(H \to \gamma\gamma)$. Similarly, the blue (dotted), 
green (dashed) and red (solid) lines represent, respectively, 
the 20\%, 10\% and 5\% limits. It is interesting to see that  
even the weakest limits displayed here are comparable to those that 
may be obtained from\footnote{Note that the analysis of 
Ref.~\cite{Baur:2004uw} did not take into consideration all the 
experimental effects and, thus, the sensitivity projected therein is 
likely to suffer further degradation.}, 
for example, the measurement of $t \bar t \gamma$
production with 30 fb$^{-1}$ of data from the LHC operating 
at a centre-of-mass energy of 14 TeV~\cite{Baur:2004uw}.

Note that, in contrast to the case for the gluon coupling, 
the constraints on the anomalous magnetic moment correspond to a much 
lower energy scale, well within the energy reach of the LHC. 
Hence it would be possible to draw definitive conclusions about the 
existence of such anomalous moments, once sufficient data has been 
accumulated.


\section{Discussions and Summary}
\label{sec:summary}

Higgs production at the LHC is dominated by gluon fusion through a top
loop. A precise measurement of the Higgs production cross-section
can, thus, be used to probe possible non-SM contributions to the $ttg$
vertex. Similarly, a measurement of the decay width of the Higgs into a
$\gamma \gamma$ final state can constrain anomalous $t t \gamma$
couplings. This assumes particular significance in the light of a
possible sighting of the Higgs by both the ATLAS and the CMS
collaborations, and the fact that the reported excess, while somewhat 
larger than that expected for a SM Higgs, is not inconsistent with such 
a hypothesis.

Parameterizing deviations of the said vertices in terms of higher-order
operators in an effective field theory, we have performed such a
study. We find that if the Higgs cross-section can be measured even to 
an accuracy of 20\%, new physics giving rise to non-standard $ttg$
couplings can be ruled out upto an energy scale of nearly 20 TeV for a
Higgs boson of mass 115--130 GeV. This would constitute an improvement
over direct constraints from $t \bar t$ production at the
LHC~\cite{Choudhury:2009wd} or even a polarized linear
collider~\cite{Rindani:1999gd,Rizzo:1994tu}. In fact, these limits
would be better than those obtained from similar (loop-induced)
indirect measurements such as $b \to s \gamma$ or the ratio $\Gamma(Z
\to b \bar b)/ \Gamma(Z \to {\rm hadrons})$~\cite{Martinez:1996cy}.
Indeed, the accuracy reach seems to be comparable to the sizes of the
anomalous chromomagnetic moment expected in a large class of
models~\cite{top_chromo_models}.

In the spirit of effective theories, we have retained terms only up to
${\cal O}({\cal A})$ or, equivalently, upto ${\cal O}(\Lambda^{-2})$
in our calculation of the cross-section. At this point, let us
consider the inclusion of the ${\cal O}({\cal A}^2)$ terms.  
As in the case of the ${\cal O}({\cal A})$ terms, equal 
and opposite logarithmic divergences are encountered for each of the 
two subsets Fig.\ref{fig:ttg_diagrams}($a,b$) and 
Fig.\ref{fig:ttg_diagrams}($c,d$).
In other words, once again, although the chromomagnetic term, on its 
own, leads to a divergent contribution, the inclusion of the full
$SU(2)_L \times U(1)_Y$ invariant term leads to a cancellation of
the divergences. Thus, at the end of the day, the ${\cal O}({\cal A}^2)$ 
contribution seems a meaningful one. Indeed, the inclusion of this 
contribution would lead to a stronger constraint on ${\cal A}$.  
We, nonetheless, omit this term while extracting limits, as we have not 
included terms higher order in $\Lambda^{-1}$ in our 
effective Lagrangian.

At this stage, we must consider possible higher-order corrections to
the Higgs production process, both within the SM and in the extended 
theory. At first sight, it seems that, with a non-renormalizable term 
in the Lagrangian, this is fraught with danger. However, experience 
with higher order calculations of the $gg \to H$ production in
the SM shows that it need not be so. NLO corrections to 
$\sigma(gg \to H)$ have been calculated in the 
Standard Model~\cite{higgs_nnlo_large_mt} using the effective field 
theory approach by introducing a term of the form 
$\lambda_{\rm eff} \, F_a^{\mu \nu} \, F_{a\mu \nu} \, H$ 
in the Lagrangian, with the coupling strength $\lambda_{\rm eff}$
being determined in terms of the one-loop calculation of this vertex 
within the SM. The difference between the infinite-$\mt$ approximation 
that such a treatment implies and the full calculation~\cite{higgs_nnlo} 
is found to be relatively small. The reason is not far to seek. 
The functional dependence on $\mt$ is a slow one and,
thus, the numerical difference caused by the $\mt \to \infty$
approximation is a small one. 
For the additional contribution due to the top chromomagnetic moment,
the story is similar, though not exactly the same. 
As eq.(\ref{eq:I}) shows, the ${\cal O}({\cal A})$ term can be 
expressed as ${\cal A} \, \mt \, f(\mt / \mh)$ where $f(x)$ is a very 
slowly varying function. Naively, the presence of the 
${\cal A} \, \mt$--factor seems to indicate that the 
infinite-$\mt$ limit is inapplicable here. However, as 
${\cal A} \sim 1/v\Lambda = 1/(v \zeta \mt)$, this apparent dependence 
on $\mt$ is only a superficial one\footnote{With $\Lambda$ being the 
  cut-off scale, all momentum integrals (and masses) have to be 
  limited to $\Lambda$ or below. In other words, the infinite-$\mt$ 
  limit can only be taken with $\mt / \Lambda$ being finite.}. 
With the remaining dependence on $\mt / \mh$ being a very slow one, 
the anomalous contribution can be subsumed in a suitably rescaled 
$\lambda_{\rm eff}$. It is, thus, quite apparent that the NLO K-factor 
in this theory would be very similar to that within the SM. 
In other words, the {\em ${\cal A}$-induced scaling} of the 
$g g \to H$ cross-section is expected to be largely insensitive to 
higher order corrections\footnote{While we have argued this for the 
  ${\cal O}({\cal A})$ term, clearly it holds as well for the 
  ${\cal O}({\cal A}^2)$ term.}.

Higher-order QCD corrections are not the only uncertainties plaguing 
the cross-section calculation. The choice of the parton distributions
as well as the factorization scale, together, have significant
uncertainties associated with them. The estimates vary, ranging from 
$\sim 10\%$~\cite{Baglio:2010um} to $\sim 20\%$~\cite{Dittmaier:2011ti}. 
This is almost irreducible in the present context and cannot
be entirely circumvented in the absence of still higher order
calculations. Indeed, this uncertainty is applicable to any effort to
establish this resonance as the SM Higgs. Comparison of the production
cross-sections across modes is expected to reduce this uncertainty
to some extent but not entirely.
Note, though, that bounds obtained from $t \bar t$ production
etc. would also be plagued by similar uncertainties. At an $e^+e^-$
collider, ratios such as $\sigma(t \bar t + {\rm jet}) / \sigma(t \bar
t)$ could be expected to give additional information. However, the
efficacy of this observable is not clear in the context of the LHC
both on account of the reduced statistics that the observation of an
additional jet entails, and also due to the uncertainties in the very
defintion of such semi-inclusive cross-sections. In view of this, it
is quite apparent that the bounds advocated here should be treated as
complementary to direct ones.

As far as top-photon interactions are concerned, the comparison of the 
Higgs signal in the diphoton mode with that in the four-lepton channel 
(with signals in both having been reported) would, for the same range 
of Higgs masses, rule out new physics contributions to the 
$t t \gamma$ vertex upto more than 5 TeV. This energy regime is also 
within reach of the LHC, and hence, it should be
possible to detect direct signals of any new physics that is involved
as well. Once again, for reasons exactly analogous to those expressed
earlier, the QCD corrections are almost identical for the SM decay and
the decay in this extended model. Furthermore, this comparison of the
modes almost entirely frees one from the aforementioned uncertainties
due to the choice of the parton distributions and the factorization
scale.

It should be appreciated that the bounds were obtained starting with
effective operators invariant under the full SM gauge symmetry. Were
we to consider only (chromo-)magnetic dipole couplings in
isolation---{\em i.e.}  admit terms that respect only 
$SU(3)_C \times U(1)_{\rm em}$---the corresponding contributions to 
the Higgs partial widths would have received logarithmic enhancements 
and the consequent bounds would have been even stronger.

Understandably, (chromo-)electric dipole moments cannot be constrained 
well using these observables. However, once sufficient data is 
accumulated to permit determination of the $CP$ properties of the 
putative resonance, even this would be possible. Indeed, were it to be 
established to be a pseudoscalar, it would be a challenging task to 
establish such large cross-sections in any given model, and 
effective operators such as those we have considered could provide a 
guideline for this task.


\textbf{Acknowledgements:} It is a pleasure to thank 
C\'{e}line Degrande, Abdelhak Djouadi and Michael Spira for critical 
comments. PS thanks CSIR, India for financial assistance under grant
09/045(0736)/2008-EMR-I.


\section*{Appendix}
\label{sec:appendix}

The diagrams of Fig.\ref{fig:ttg_diagrams}$(a,b)$ give us, 
for the effective $g g H$ vertex, 
\begin{align}
{\cal M}_1^{\mu\nu}
\quad &= \quad
\bigg( g_s^2 \dfrac{\mt}{v} \bigg) \;
Tr[T_a T_b] \;
\int \dfrac{d^4k}{(2\pi)^4} \,
\dfrac
{Tr[(\kay + \pone + \mt) \, \Gamma^{\mu}_1 \, (\kay + \mt) \, \Gamma^{\nu}_2 \, (\kay - \ptwo + \mt)]}
{[(k + p_1)^2 - \mt^2][k^2 - \mt^2][(k-p_2)^2 - \mt^2]}
\\[6ex]
{\cal M}_2^{\mu\nu}
\quad &= \quad
-\bigg( g_s^2 \dfrac{\mt}{v} \bigg) \;
Tr[T_b T_a] \;
\int \dfrac{d^4k}{(2\pi)^4} \,
\dfrac
{Tr[(\kay - \ptwo - \mt) \, \Gamma^{\nu}_2 \, (\kay - \mt) \, \Gamma^{\mu}_1 \, (\kay + \pone - \mt)]}
{[(k - p_2)^2 - \mt^2][k^2 - \mt^2][(k + p_1)^2 - \mt^2]} \ .
\end{align}

Similarly,  those in Fig.\ref{fig:ttg_diagrams}$(c,d)$ lead to 
\begin{align}
{\cal M}_3^{\mu\nu}
\quad &= \quad
-g_s^2 \;
Tr[T_a T_b] \;
\int \dfrac{d^4k}{(2\pi)^4} \,
\dfrac
{Tr[ \widetilde\Gamma^{\mu}_1 \, (\kay + \mt) \, \Gamma^{\nu}_2 \, (\kay - \ptwo + \mt)]}
{[k^2 - \mt^2][(k - p_2)^2 - \mt^2]}
\\[4ex]
{\cal M}_4^{\mu\nu}
\quad &= \quad
-g_s^2 \;
Tr[T_b T_a] \;
\int \dfrac{d^4k}{(2\pi)^4} \,
\dfrac
{Tr[ \widetilde\Gamma^{\nu}_2 \, (\kay + \mt) \, \Gamma^{\mu}_1 \, (\kay -\pone + \mt)]}
{[k^2 - \mt^2][(k - p_1)^2 - \mt^2]} \ ,
\end{align}
where
\begin{align}
\Gamma^{\mu}_1
\quad &= \quad
\gamma^{\mu} + \sqrt{2} \, i \, {\cal B} \, v \, \sigma^{\mu\alpha} \, p_{1\alpha} 
\mspace{30mu} , \mspace{30mu}
\widetilde\Gamma^{\mu}_1
\quad = \quad
\sqrt{2} \, i \, {\cal B} \, \sigma^{\mu\alpha} \, p_{1\alpha} 
\\[2ex]
\Gamma^{\nu}_2
\quad &= \quad
\gamma^{\nu} + \sqrt{2} \, i \, {\cal B} \, v \, \sigma^{\nu\beta} \, p_{2\beta} 
\mspace{30mu} , \mspace{30mu}
\widetilde\Gamma^{\nu}_2
\quad = \quad
\sqrt{2} \, i \, {\cal B} \, \sigma^{\nu\beta}p_{2\beta}  \ .
\end{align}
with ${\cal B} = Re({\cal A}) + i Im({\cal A}) \, \gamma_5 $.

Whereas ${\cal M}_{1,2}^{\mu\nu}$ have apparent linear divergences,
for ${\cal M}_{3,4}^{\mu\nu}$ the naive degree of divergence is
quadratic. For on-shell gluons, terms proportional to 
$p_1^2$ and $p_2^2$ vanish identically. Since the gluon couples to a 
conserved current, terms proportional to $p_1^{\mu}$ or $p_2^{\nu}$ 
vanish as well. Consequently, the quadratic and linear divergences 
disappear, leaving behind terms that are either finite or, at worst,
logarithmically divergent. The divergences can be regularized using 
any gauge-invariant prescription such as dimensional regularization.
On summing all the contributing amplitudes and performimg 
dimensional regularization, one obtains a finite and gauge-invariant 
form for the vertex function as a series in ${\cal A}$, namely
\begin{equation}
{\cal M}^{\mu\nu}
= i \bigg( \dfrac{g_s^2}{4 \pi^2} \bigg) \delta_{ab} \,
\left [\, 
C_0 I \,+\, C_1 J \,+\, C_2 \left(J - \dfrac{1}{2}\right) \,
\right] 
\label{eq:full_amplitude}
\end{equation}

\begin{align}
C_0 
\quad &= \quad
\dfrac{1}{v} \,
\bigg( \dfrac{\mh^2}{2} g^{\mu\nu} - p_1^{\nu} p_2^{\mu} \bigg) \\[2ex]
C_1
\quad &= \quad
4 \mt \,
\bigg[
\dfrac{Re({\cal A})}{\sqrt{2}} \, 
\bigg( \dfrac{\mh^2}{2} g^{\mu\nu} - p_1^{\nu}p_2^{\mu} \bigg)
\,+\,
\dfrac{Im({\cal A})}{\sqrt{2}} \, \bigg( \epsilon^{\mu\alpha\beta\nu} \, p_{1\alpha} p_{2\beta} \bigg)
\bigg] \\[2ex]
C_2
\quad &= \quad
4 \mt^2 v
\left[ 
Re({\cal A}^2) \, \bigg( \dfrac{\mh^2}{2} g^{\mu\nu} - p_1^{\nu}p_2^{\mu} \bigg)
\,+\,
Im({\cal A}^2) \, \epsilon^{\mu\alpha\beta\nu} p_{1\alpha} p_{2\beta}
\right]
   \label{eq:coeffs}
\end{align}

Defining $w = \dfrac{\mh^2}{m_t^2}$, we have for the integrals
\begin{align}
I(w)
 &= 
\int_0^1 dx
\int_0^{1-x} dy \;
\left( \dfrac{1-4xy}{1 - wxy} \right) \nonumber \\[2ex]
 &= 
\dfrac{2}{w} 
\,-\,
\bigg( \dfrac{4 - w}{w^2} \bigg)
\bigg[
{\rm Li}_2 \bigg( \dfrac{2 w}{ w + \sqrt{w^2 - 4w}} \bigg)
+
{\rm Li}_2 \bigg( \dfrac{2 w}{ w - \sqrt{w^2 - 4w}} \bigg)
\bigg]
   \label{eq:I}
\\[6ex]
J(w)
&=
\int_0^1 dx
\int_0^{1-x} dy
\bigg[
\log \bigg( 1 - wxy \bigg) + 2
\bigg] \nonumber \\[2ex]
 &= \;
-\dfrac{1}{2} 
\,+\,
\sqrt{\dfrac{4-w}{w}} \, \tan^{-1}\sqrt{\dfrac{w}{4-w}}  
\,+\,
\dfrac{1}{w}
\bigg[
{\rm Li}_2 \bigg( \dfrac{2 w}{ w + \sqrt{w^2 - 4w}} \bigg)
+
{\rm Li}_2 \bigg( \dfrac{2 w}{ w - \sqrt{w^2 - 4w}} \bigg)
\bigg] 
   \label{eq:J}
\end{align}
where ${\rm Li}_2(x)$ is the dilogarithm or Spence function.


\newpage

\begin{small}

\end{small}


\begin{thebibliography}{99}

\bibitem{Barate:2003sz}
  R.~Barate {\it et al.}  [LEP Working Group for Higgs boson searches],
  Phys.\ Lett.\  B {\bf 565}, 61 (2003)
  [arXiv:hep-ex/0306033].

\bibitem{ATLAS_Dec2011}
  G.~Aad {\it et al.}  [ATLAS Collaboration],
  Phys.\ Lett.\ B {\bf 710}, 49 (2012)
  [arXiv:1202.1408 [hep-ex]].


\bibitem{CMS_Dec2011}
  S.~Chatrchyan {\it et al.}  [CMS Collaboration],
  Phys.\ Lett.\ B {\bf 710}, 26 (2012)
  [arXiv:1202.1488 [hep-ex]].

\bibitem{CDFandD0:2011aa} 
  TEVNPH (Tevatron New Phenomena and Higgs Working Group) and CDF and D0 Collaborations,
  arXiv:1107.5518 [hep-ex].


\bibitem{pheno_refs_1}
  L.~J.~Hall, D.~Pinner and J.~T.~Ruderman,
  JHEP {\bf 1204}, 131 (2012)
  [arXiv:1112.2703 [hep-ph]];\\
  S.~Heinemeyer, O.~Stal and G.~Weiglein,
  Phys.\ Lett.\ B {\bf 710}, 201 (2012)
  [arXiv:1112.3026 [hep-ph]];\\
  A.~Arbey {\it et al.},
  Phys Lett. {\bf B708}, 162  (2012) [arXiv:1112.3028 [hep-ph]];\\
  M.~Carena {\it et al.},
  JHEP {\bf 1203}, 014 (2012)
  [arXiv:1112.3336 [hep-ph]];\\
  U.~Ellwanger,
  JHEP {\bf 1203}, 044 (2012)
  [arXiv:1112.3548 [hep-ph]];\\
  M.~Kadastik {\it et al.},
  JHEP {\bf 1205}, 061 (2012)
  [arXiv:1112.3647 [hep-ph]];\\
  J.~Cao, Z.~Heng, D.~Li and J.~M.~Yang,
  Phys.\ Lett.\ B {\bf 710}, 665 (2012)
  [arXiv:1112.4391 [hep-ph]];\\
  B.~Batell, S.~Gori and L.~T.~Wang,
  arXiv:1112.5180 [hep-ph];\\
  J.~F.~Gunion, Y.~Jiang and S.~Kraml,
  Phys.\ Lett.\ B {\bf 710}, 454 (2012)
  [arXiv:1201.0982 [hep-ph]];\\
  P.~Fileviez Perez,
  Phys.\ Lett.\ B {\bf 711}, 353 (2012)
  [arXiv:1201.1501 [hep-ph]];\\
  S.~F.~King, M.~Muhlleitner and R.~Nevzorov,
  Nucl.\ Phys.\ B {\bf 860}, 207 (2012)
  [arXiv:1201.2671 [hep-ph]].

\bibitem{pheno_refs_2}
  D.~Albornoz Vasquez, G.~Belanger, R.~M.~Godbole and A.~Pukhov,
  arXiv:1112.2200 [hep-ph];\\
  A.~Arbey, M.~Battaglia and F.~Mahmoudi,
  Eur.\ Phys.\ J.\ C {\bf 72}, 1906 (2012)
  [arXiv:1112.3032 [hep-ph]];\\
  C.~Cheung and Y.~Nomura,
  arXiv:1112.3043 [hep-ph];\\
  N.~Karagiannakis, G.~Lazarides and C.~Pallis,
  arXiv:1201.2111 [hep-ph].

\bibitem{pheno_refs_3}
  G.~Guo, B.~Ren and X.~G.~He,
  arXiv:1112.3188 [hep-ph];\\
  G.~Burdman, C.~Haluch and R.~Matheus,
  Phys.\ Rev.\ D {\bf 85}, 095016 (2012)
  [arXiv:1112.3961 [hep-ph]];\\
  K.~Cheung and T.~-C.~Yuan,
  Phys.\ Rev.\ Lett.\  {\bf 108}, 141602 (2012)
  [arXiv:1112.4146 [hep-ph]];\\
  F.~Goertz, U.~Haisch and M.~Neubert,
  Phys.\ Lett.\ B {\bf 713}, 23 (2012)
  [arXiv:1112.5099 [hep-ph]];\\
  X.~-G.~He, B.~Ren and J.~Tandean,
  Phys.\ Rev.\ D {\bf 85}, 093019 (2012)
  [arXiv:1112.6364 [hep-ph]].

\bibitem{LEPEWWG}
The LEP Electroweak Working Group, http://lepewwg.web.cern.ch/LEPEWWG

\bibitem{DyEWSB}
  R.~S.~Chivukula, M.~Narain and J.~Womersley,
  pages 1258-1264 of Review of Particle Physics,\\ 
  (C.~Amsler {\it et al.}  [Particle Data Group]),
  Phys.\ Lett.\  B {\bf 667}, 1 (2008); \\
  G.~Bhattacharyya,
  Rept.\ Prog.\ Phys.\  {\bf 74}, 026201 (2011)
  [arXiv:0910.5095 [hep-ph]]; \\
  G.~Isidori,
  PoS C {\bf D09}, 073 (2009)
  [arXiv:0911.3219 [hep-ph]].

\bibitem{tt_condensate}
  C.~T.~Hill,
  Phys.\ Lett.\  B {\bf 266}, 419 (1991); \\
  C.~T.~Hill,
  Phys.\ Lett.\  B {\bf 345}, 483 (1995)
  [arXiv:hep-ph/9411426].

\bibitem{lh0}
  N.~Arkani-Hamed, A.~G.~Cohen and H.~Georgi,
  Phys.\ Lett.\  B {\bf 513}, 232 (2001)
  [arXiv:hep-ph/0105239]; \\
\textit{For reviews, see, for example,} \\
  M.~Schmaltz and D.~Tucker-Smith,
  Ann.\ Rev.\ Nucl.\ Part.\ Sci.\  {\bf 55}, 229 (2005)
  [arXiv:hep-ph/0502182]; \\
  M.~Perelstein,
  Prog.\ Part.\ Nucl.\ Phys.\  {\bf 58}, 247 (2007)
  [arXiv:hep-ph/0512128], \\
\textit{and references therein}.

\bibitem{acd_ued}
  T.~Appelquist {\it et al.},
  Phys.\ Rev.\  D {\bf 64}, 035002 (2001)
  [arXiv:hep-ph/0012100].

\bibitem{ued_others}
  I.~Antoniadis,
  Phys.\ Lett.\  B {\bf 246}, 377 (1990); \\
  N.~Arkani-Hamed and M.~Schmaltz,
  Phys.\ Rev.\  D {\bf 61}, 033005 (2000)
  [arXiv:hep-ph/9903417].

\bibitem{Barbieri_ed}
  R.~Barbieri, L.~J.~Hall and Y.~Nomura,
  Phys.\ Rev.\  D {\bf 63}, 105007 (2001)
  [arXiv:hep-ph/0011311]; \\
  G.~Cacciapaglia, M.~Cirelli and G.~Cristadoro,
  Nucl.\ Phys.\  B {\bf 634}, 230 (2002)
  [arXiv:hep-ph/0111288].

\bibitem{diquark}
  J.~L.~Hewett and T.~G.~Rizzo,
  Phys.\ Rept.\  {\bf 183}, 193 (1989); \\
  G.~Bhattacharyya {\it et al.},
  Phys.\ Lett.\  B {\bf 355}, 193 (1995)
  [arXiv:hep-ph/9504314]; \\
  T.~Han, I.~Lewis and Z.~Liu,
  JHEP {\bf 1012}, 085 (2010)
  [arXiv:1010.4309 [hep-ph]].

\bibitem{eff_terms} 
  J.~A.~Aguilar-Saavedra,
  Nucl.\ Phys.\  B {\bf 812}, 181 (2009)
  [arXiv:0811.3842 [hep-ph]]; \\
  W.~Buchmuller and D.~Wyler,
  Nucl.\ Phys.\  B {\bf 268}, 621 (1986); \\
  C.~Arzt, M.~B.~Einhorn and J.~Wudka,
  Nucl.\ Phys.\  B {\bf 433}, 41 (1995)
  [arXiv:hep-ph/9405214].

\bibitem{previous}
  D.~Atwood, A.~Aeppli and A.~Soni,
  Phys.\ Rev.\ Lett.\  {\bf 69}, 2754 (1992); \\
  T.~G.~Rizzo,
  arXiv:hep-ph/9407366; \\
  D.~Atwood, A.~Kagan and T.~G.~Rizzo,
  Phys.\ Rev.\  D {\bf 52}, 6264 (1995)
  [arXiv:hep-ph/9407408]; \\
  P.~Haberl, O.~Nachtmann and A.~Wilch,
  Phys.\ Rev.\  D {\bf 53}, 4875 (1996)
  [arXiv:hep-ph/9505409]; \\
  K.~M.~Cheung,
  Phys.\ Rev.\  D {\bf 53}, 3604 (1996)
  [arXiv:hep-ph/9511260]; \\
  T.~G.~Rizzo,
  [arXiv:hep-ph/9609311]; \\
  S.~Y.~Choi, C.~S.~Kim and J.~Lee,
  Phys.\ Lett.\  B {\bf 415}, 67 (1997)
  [arXiv:hep-ph/9706379]; \\
  B.~Grzadkowski, B.~Lampe and K.~J.~Abraham,
  Phys.\ Lett.\  B {\bf 415}, 193 (1997)
  [arXiv:hep-ph/9706489]; \\
  B.~Lampe,
  Phys.\ Lett.\  B {\bf 415}, 63 (1997)
  [arXiv:hep-ph/9709493]; \\
  H.~Y.~Zhou,
  Phys.\ Rev.\  D {\bf 58}, 114002 (1998)
  [arXiv:hep-ph/9805358]; \\
  K.~I.~Hikasa {\it et al.},
  Phys.\ Rev.\  D {\bf 58}, 114003 (1998)
  [arXiv:hep-ph/9806401]; \\
  K.~Ohkuma,
  arXiv:hep-ph/0105117; \\
  J.~Sjolin,
  J.\ Phys.\ G {\bf 29}, 543 (2003); \\
  D.~Atwood, S.~Bar-Shalom, G.~Eilam and A.~Soni,
  Phys.\ Rept.\  {\bf 347}, 1 (2001)
  [arXiv:hep-ph/0006032]; \\
Z.~Dong, G.~Durieux, J.~M.~Gerard, T.~Han and F.~Maltoni,
arXiv:1107.3805 [hep-ph].

\bibitem{T_odd_corr}
  O.~Antipin and G.~Valencia,
  Phys.\ Rev.\  D {\bf 79}, 013013 (2009)
  [arXiv:0807.1295 [hep-ph]];\\
  S.~K.~Gupta, A.~S.~Mete and G.~Valencia,
  Phys.\ Rev.\  D {\bf 80}, 034013 (2009)
  [arXiv:0905.1074 [hep-ph]].

\bibitem{Choudhury:2009wd}
D.~Choudhury and P.~Saha,
Pramana {\bf 77}, 1079 (2011)
[arXiv:0911.5016 [hep-ph]];\\
P.~Saha,
arXiv:1010.1931 [hep-ph].

\bibitem{Rindani:1999gd}
  S.~D.~Rindani and M.~M.~Tung,
  Eur.\ Phys.\ J.\  C {\bf 11}, 485 (1999)
  [arXiv:hep-ph/9904319].

\bibitem{Atwood:1991ka}
  D.~Atwood and A.~Soni,
  Phys.\ Rev.\  D {\bf 45}, 2405 (1992);\\
  S.~Y.~Choi and K.~Hagiwara,
  Phys.\ Lett.\  B {\bf 359}, 369 (1995)
  [arXiv:hep-ph/9506430].

\bibitem{Rizzo:1994tu}
  T.~G.~Rizzo,
  Phys.\ Rev.\  D {\bf 50}, 4478 (1994)
  [arXiv:hep-ph/9405391];
  arXiv:hep-ph/9605361.

\bibitem{Martinez:1996cy}
  R.~Martinez and J.~A.~Rodriguez,
  Phys.\ Rev.\  D {\bf 55}, 3212 (1997)
  [arXiv:hep-ph/9612438];\\
  R.~Martinez and J.~A.~Rodriguez,
  Phys.\ Rev.\  D {\bf 60}, 077504 (1999)
  [arXiv:hep-ph/9707453].

\bibitem{hasenfratz}
  P.~Hasenfratz and J.~Nager,
  Z.\ Phys.\ C {\bf 37}, 477 (1988);\\
  S.~R.~Choudhury and Mamta,
  Int.\ J.\ Mod.\ Phys.\ A {\bf 12}, 1847 (1997).

\bibitem{Baur:2004uw} 
  U.~Baur, A.~Juste, L.~H.~Orr and D.~Rainwater,
  Phys.\ Rev.\ D {\bf 71}, 054013 (2005)
  [hep-ph/0412021].

\bibitem{top_chromo_models}
  R.~Martinez and J.~A.~Rodriguez,
  Phys.\ Rev.\  D {\bf 65}, 057301 (2002)
  [arXiv:hep-ph/0109109];\\
  R.~Martinez, M.~A.~Perez and N.~Poveda,
  Eur.\ Phys.\ J.\  C {\bf 53}, 221 (2008)
  [arXiv:hep-ph/0701098];\\
  L.~Ding and C.~X.~Yue,
  Commun.\ Theor.\ Phys.\  {\bf 50}, 441 (2008)
  [arXiv:0801.1880 [hep-ph]];\\
  Q.~H.~Cao {\it et al.},
  Phys.\ Rev.\  D {\bf 79}, 015004 (2009)
  [arXiv:0801.2998 [hep-ph]].

\bibitem{higgs_nnlo_large_mt}
 S. Dawson, 
  Nucl. Phys. {\bf B359}, 283 (1991);\\
%
  A. Djouadi, M. Spira, and P. M. Zerwas, 
  Phys. Lett. {\bf B264}, 440 (1991);\\
%
  R.~V.~Harlander and W.~B.~Kilgore,
  Phys.\ Rev.\ Lett.\  {\bf 88}, 201801 (2002)
  [arXiv:hep-ph/0201206];\\
%
  C.~Anastasiou and K.~Melnikov,
  Nucl.\ Phys.\  B {\bf 646}, 220 (2002)
  [arXiv:hep-ph/0207004];\\
%
  V.~Ravindran, J.~Smith and W.~L.~van Neerven,
  Nucl.\ Phys.\  B {\bf 665}, 325 (2003)
  [arXiv:hep-ph/0302135].

\bibitem{higgs_nnlo}
  D.~Graudenz, M.~Spira and P.~M.~Zerwas,
  Phys.\ Rev.\ Lett.\  {\bf 70}, 1372 (1993);\\
  M.~Spira, A.~Djouadi, D.~Graudenz and P.~M.~Zerwas,
  Nucl.\ Phys.\ B {\bf 453}, 17 (1995)
  [hep-ph/9504378].

\bibitem{Baglio:2010um}
  J.~Baglio and A.~Djouadi,
  JHEP {\bf 1010}, 064 (2010)
  [arXiv:1003.4266 [hep-ph]].

\bibitem{Dittmaier:2011ti}
  S.~Dittmaier {\it et al.}  [LHC Higgs Cross Section Working Group],
  arXiv:1101.0593 [hep-ph].

\end{thebibliography}
\end{document}